\RequirePackage{etex}
\documentclass[5p,times]{cas-dc}
\usepackage[sort, compress, numbers]{natbib}

\usepackage{lastpage}
\usepackage{fancyhdr}

\fancypagestyle{plain}{
    \fancyhf{}
    \fancyfoot[R]{Page \thepage\ of \pageref{LastPage}}
    
}
\usepackage{epsfig}
\usepackage{graphicx}
\usepackage{xcolor}
\usepackage{pgfplots}
\pgfplotsset{compat=1.17}

\usepackage{amssymb}
\usepackage{amsmath}
\usepackage{multirow}
\usepackage{amsfonts}
\usepackage{dashbox}

\usepackage{changepage}
\usepackage{array}
\usepackage{makecell}
\usepackage{hyperref} 
\usepackage{bookmark} 
\usepackage{pifont}
\usepackage{algorithmic}
\usepackage{textcomp}
\usepackage{verbatim}
\usepackage{tcolorbox}
\usepackage{longtable}
\usepackage{multirow}
\usepackage{framed} 
\usepackage{tabularx} 
\usepackage{booktabs} 
\usepackage{ragged2e} 
\usepackage{float} 

\hypersetup{
    colorlinks=true,          
    linkcolor=blue,           
    citecolor=blue,           
    filecolor=magenta,        
    urlcolor=blue,            
}
\usepackage[justification=centering]{caption}
\usepackage{circledsteps}

\def\BibTeX{{\text{B\kern-.05em{\sc i\kern-.025em b}\kern-.08em\TeX}}}

\begin{document}

\shorttitle{LLM Access Shield: Domain-Specific LLM Framework for Privacy Policy Compliance}

\shortauthors{Wang et~al.}

\title [mode = title]{LLM Access Shield: Domain-Specific LLM Framework  for Privacy Policy Compliance}                      
\author{Yu Wang}
\author{Cailing Cai}
\author{Zhihua Xiao}
\author{Peifung E. Lam}[orcid=0009-0007-3991-7683]
\cormark[1]
\ead{ericlam@astri.org}

\affiliation{organization={Hong Kong Applied Science and Technology Research Institute (ASTRI)},
    country={Hong Kong SAR}}

\cortext[cor1]{Corresponding author}

\begin{abstract}
Large language models (LLMs) are increasingly applied in fields such as finance, education, and governance due to their ability to generate human-like text and adapt to specialized tasks. However, their widespread adoption raises critical concerns about data privacy and security, including the risk of sensitive data exposure. 

In this paper, we propose a security framework to enforce policy compliance and mitigate risks in LLM interactions. Our approach introduces three key innovations: 
(i) LLM-based policy enforcement: a customizable mechanism that enhances domain-specific detection of sensitive data.
(ii) Dynamic policy customization: real-time policy adaptation and enforcement during user-LLM interactions to ensure compliance with evolving security requirements. 
(iii) Sensitive data anonymization: a format-preserving encryption technique that protects sensitive information while maintaining contextual integrity.
Experimental results demonstrate that our framework effectively mitigates security risks while preserving the functional accuracy of LLM-driven tasks. 
\end{abstract}


\maketitle

\section{Introduction}
\label{sect:introduction}


Large language models (LLMs), such as Microsoft’s Copilot, OpenAI’s GPT, and Google’s Gemini, have substantially advanced a wide range of applications, including text summarization, content generation, and software development. 
Despite their impressive capabilities, these LLM services pose significant privacy risks. User prompts, which often contain sensitive personal or organizational information, are transmitted to third-party servers, where they may be vulnerable to data breaches, unauthorized access, or inference attacks. 
Recent work by Chu et al. \citep{DBLP:conf/emnlp/ChuS0024} demonstrated that adversaries could exploit GPT models to extract private information through carefully crafted prompts, emphasizing the critical importance of privacy protection in LLM interactions.

Various privacy-preserving techniques have been developed for LLMs, including cryptography-based and perturbation-based methods. Cryptography-based approaches include
homomorphic encryption (HE) \citep{elgamal1985HE,paillier1999HE,gentry2009fully,BV11HE,fan2012HE,ckks2017HE,21-LM-HE-EUROCRYPT}, which enables computations to be performed directly on encrypted data,
and secure multi-party computation (MPC) \citep{cramer2015mpc,asharov2013mpc}, which facilitates collaborative computation among multiple parties without revealing their private inputs.
Both approaches offer solid theoretical bases and are widely used in privacy-preserving machine learning.  
For example, {\it Cryptonets} \citep{16-RNKKM-Cryptonets-HE-ICML} and {\it Iron}\citep{22-HLC+Iron} have demonstrated the feasibility of neural network inference on encrypted data. 
However, the computational cost of {\it HE}-based  \citep{16-RNKKM-Cryptonets-HE-ICML,22-CBH+THE-X-PPInferenceHE,22-HLC+Iron,23-LHG+Bumblebee-HE,24-ZBD+HE,23-LL-LLM-HE}  and {\it MPC}-based \citep{DBLP:22-WSX+ISPASS-MPC,23-HLL+Ciphergpt-MPC,23-DGG+East,23-AFAS-Privformer-MPC-EuroSP,23-DLZ+Puma,23-GJM+2PC-FSS} schemes is excessively high.  
Moreover, both methods require significant modifications to the LLM infrastructure, rendering them impractical for real-time interactions with LLM.   
Perturbation-based methods, such as {\it differential privacy} \citep{22-RMJB-EWTune-DP,22-SSCZJY-JustFineTuneTwice-DP,22-WGX-DP,22-MDPSGZ-DP,23-DYCWHS-DP-Forward-ccs,23-MYHYP-Split-and-Denoise-DP,23-INS+PrivChatGPT-DP,23-TCQ+Privinfer,23-LTL-RAPT,24-HWZ+ICLR-DP,23-UHG+PrivChatGPT}, aim to protect sensitive information by adding noise to user prompts, model parameters, or outputs.  Despite their effectiveness, such methods often degrade task performance, thus limiting the practical utility of LLM responses.

In response to these challenges, several frameworks have been developed to identify and obfuscate sensitive information in user prompts, supporting privacy protection without modifying the LLM architecture.  
Notable approaches include {\it EmojiCrypt} \citep{24-LHZ-EmojiCrypt}, {\it ProSan} \citep{24-SXHTHZ-ProSan},  {\it Casper} \citep{24-CHYT-Casper}, and {\it ConfusionPrompt} \citep{24-MYYYP-ConfusionPrompt}. 
Although these frameworks mark promising advancements, 
they face challenges in achieving a practical balance between privacy, contextual integrity, and utility. 
For example, {\it EmojiCrypt} \citep{24-LHZ-EmojiCrypt}, relies on a limited symbolic vocabulary to encode sensitive information, which may result in oversimplification or loss of important semantic details. As a result, contextually critical information may be misrepresented, leading to inaccurate inferences. For instance, using an emoji to denote a product's manufacturing date could lead to incorrect attributes.  
{\it ProSan} \citep{24-SXHTHZ-ProSan}, which depends on dynamic assessments of word importance and privacy risks, may face challenges such as inadequate protection of sensitive information in ambiguous contexts, and potential degradation of task performance due to over-anonymization. Furthermore, it remains unclear whether {\it ProSan} reverts obfuscated words to their original form after response generation, leaving questions about its usability in end-to-end workflows.
To identify privacy-sensitive topics in the prompts, {\it Casper} \citep{24-CHYT-Casper} incorporates a pre-trained LLM (e.g., Llama 2, Llama 3, etc.) that operates locally within a browser extension, ensuring sensitive data does not leave the user's device. Although this approach improves data security, such models often lack the precision needed for domain-specific tasks, increasing the likelihood of errors or omissions when handling specialized content. 
Besides, deploying a pre-trained LLM locally requires massive computational resources, including high-performance GPUs, large amounts of memory, high-speed storage, high power consumption, and complex software optimizations to ensure efficient inference and stable operation. 
For {\it ConfusionPrompt} \citep{24-MYYYP-ConfusionPrompt}, utilizing multiple sub-prompts or pseudo-prompts to obfuscate sensitive information, may introduce additional computational and communication overhead. Furthermore, the scheme needs to consolidate the responses from all sub-prompts to reconstruct the final output, increasing user's query operations and server computation burden.

\begin{table*}[h!t]
    \centering
    \scriptsize
    \begin{tabular}{|c|c|c|c|c|c|}
        \hline
              & \begin{tabular}[c]{@{}c@{}}Support Model \\ Policy Customization?\end{tabular} 
              & \begin{tabular}[c]{@{}c@{}}Identify  \\ Malicious  Prompts?\end{tabular} 
              & \begin{tabular}[c]{@{}c@{}}Revert   SI \\ from  Response?\end{tabular}  
              & \begin{tabular}[c]{@{}c@{}}Method for \\ Protecting SI\end{tabular}
              & \begin{tabular}[c]{@{}c@{}}Maintain \\ Contextual  Integrity? \end{tabular}              
              \\ \hline
        
        {\it EmojiCrypt} \citep{24-LHZ-EmojiCrypt}     
        & $\times$
        & $\times$                                                                  
        & $\checkmark$
        
        & Emoji replacement 
        & $\times$                                  
        \\ \hline
        
        {\it ProSan} \citep{24-SXHTHZ-ProSan}     
        & $\times$
        & $\times$ 
        & $\times$   
        
        & \begin{tabular}[c]{@{}c@{}}Importance/privacy-score \\based replacement\end{tabular}
        & $\checkmark$
        \\ \hline
        
        {\it Casper} \citep{24-CHYT-Casper}     
        & $\times$
        & $\times$
        & $\checkmark$   
        
        & Placeholder use
        & $\checkmark$ 
        \\ \hline
        
        {\it ConfusionPrompt} \citep{24-MYYYP-ConfusionPrompt}     
        & $\times$
        & $\times$                                                                
        & $\checkmark$  
        
        & \begin{tabular}[c]{@{}c@{}}Multi pseudo-prompts\end{tabular}
        & $\times$                                   
        \\ \hline
        
        This Work 
        & \begin{tabular}[c]{@{}c@{}}$\checkmark$ (support domain  specific  \\ tasks \&  dynamic policy control)\end{tabular}                                 
        & $\checkmark$
        & $\checkmark$        
        
        & FPE encryption
        & $\checkmark$        
        \\ \hline
    \end{tabular}
    \caption{Comparison of methods for protecting user privacy, including their customization capabilities, ability to identify malicious prompts (i.e., detecting prompts that may lead to unauthorized data access, policy violations, or security risks in enterprise environments), and capability to revert sensitive information (SI) from responses. The table also includes the protection method used and whether the user query's contextual integrity is maintained after SI processing. FPE denotes Format-Preserving Encryption.}
    \label{tab:Comparison of Different Methods}
\end{table*}

Our work differs from previous approaches by focusing on the construction of customized LLMs, privacy protection, and contextual integrity of prompts and responses.  
Although existing LLMs (such as LLaMA and its variants) provide broad functionality, they lack the capabilities to enforce unique sensitivity policies. 
Nevertheless, none of the existing works \citep{24-LHZ-EmojiCrypt,24-SXHTHZ-ProSan,24-CHYT-Casper,24-MYYYP-ConfusionPrompt} offer a viable solution to address this challenge.  
Moreover, prior researches primarily address scenarios where users unintentionally disclose sensitive data when interacting with LLM services. 
In contrast,  35\% of security incidents arise from both deliberate and accidental employee actions \citep{herrera2022survey}. 
For instance, employees may unintentionally expose confidential data (e.g., financial statements) while using LLMs for tasks such as summarization, comprehension, or problem-solving \citep{vaishnav2024transparency}. 
Such actions could lead to compliance risks, cybersecurity threats, or intellectual property violations.  
Therefore, automatically detecting and protecting sensitive content is essential for the secure use of LLMs in organizational settings.

We provide a high-level comparison with related methods in Table~\ref{tab:Comparison of Different Methods} to highlight the distinctions of our approach. 
Our main contributions are summarized as follows:
\begin{enumerate}
    \item 
    We propose an end-to-end secure interaction framework between users and LLMs, termed \textbf{LLM Access Shield},
    which protects user privacy in organizational environments through a two-stage process: (a) detecting privacy risks using a domain-specific LLM, and (b) anonymizing unsafe prompts via a utility-preserving anonymization technique.

    \item
    We develop a domain-specific LLM, named \textbf{DLMS}, for sensitive information detection in user prompts. 
    Inspired by Llama-Guard, DLMS is trained via supervised fine-tuning (SFT) and supports downstream tasks, including token-level anonymization and privacy risk assessment.

    \begin{itemize}
        \item 
        We apply reinforcement fine-tuning (RFT) with an ``analyze-then-decide'' reasoning paradigm to prompt-level privacy risk assessment. This approach empowers models to generalize across diverse and complex privacy policies, facilitating dynamic compliance through test-time computation.

        \item We evaluate the generalization capabilities of both SFT and RFT models on unseen, non-taxonomy-based privacy policies. Our results show that the ``analyze-then-decide'' paradigm, powered by RFT, generalizes more effectively and allows dynamic, real-time policy compliance without retraining after deployment.

        \item 
        To improve the stability and effectiveness of RFT, we investigate curriculum learning strategies and propose mitigation techniques to address reward hacking behaviors, offering actionable insights for training robust, privacy-aware LLMs.
    \end{itemize}
    
    \item
    We adopt format-preserving encryption (FPE) as a utility-preserving anonymization method to protect sensitive entities identified by DLMS, while maintaining the semantic integrity of the original prompts.
\end{enumerate}

\begin{figure*}[t]
    \centering
    \includegraphics[width=\linewidth]{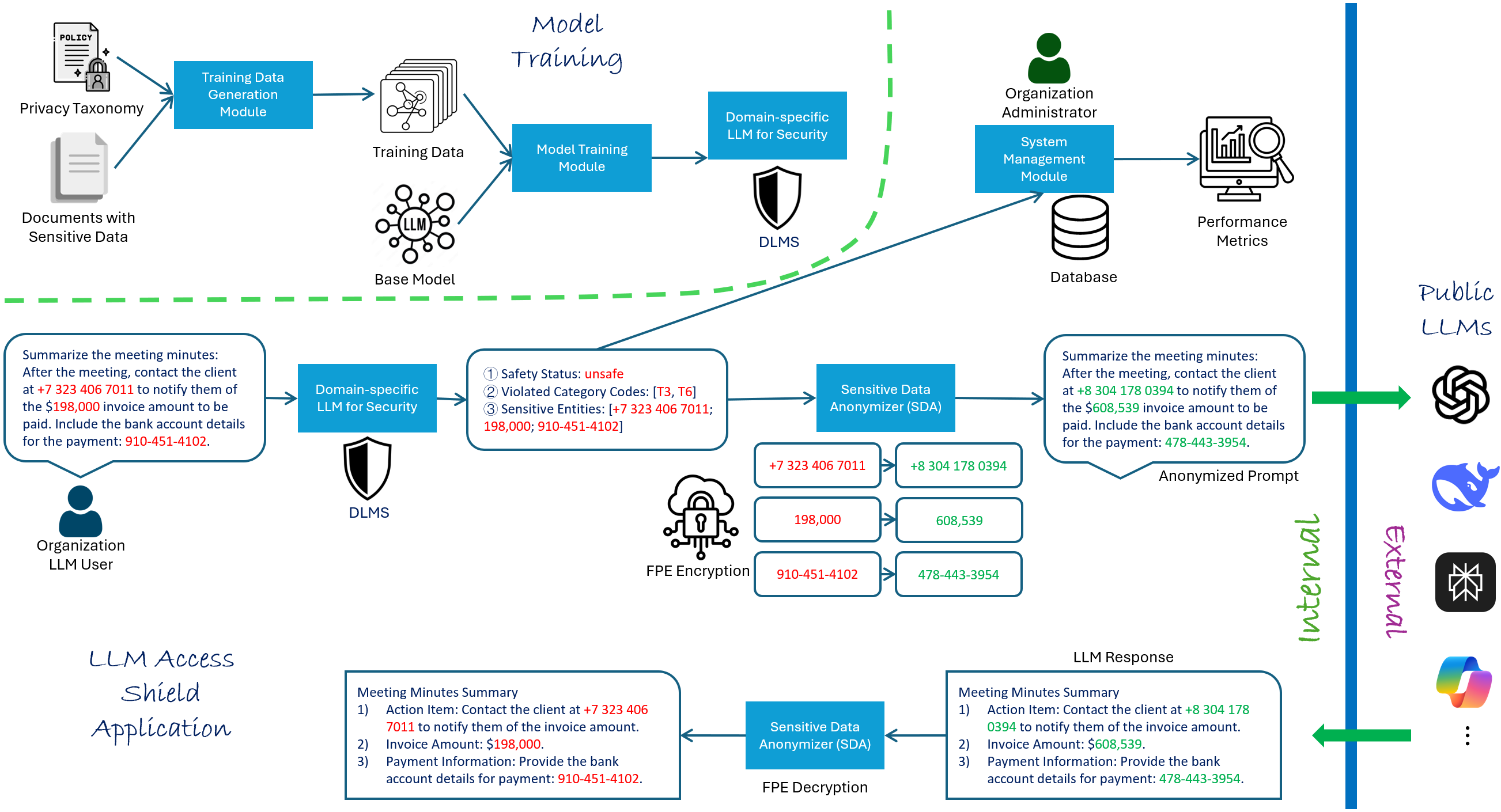}
    \caption{LLM Access Shield: System Infrastructure.}
    \label{fig:Framework}
\end{figure*}

\section{LLM Access Shield Framework}
\label{sect:framework}
In this section, we introduce LLM Access Shield, a lightweight and extensible framework designed to safeguard user privacy in high-assurance domains such as finance, law, and healthcare.

The framework requires minimal integration effort and supports continuous adaptation to evolving privacy requirements through monitoring and policy updates, making it suitable for real-world deployment.




\subsection{System Overview}
Fig. \ref{fig:Framework} illustrates the application process of the LLM Access Shield framework and the model training process of DLMS. 
The application process primarily involves the organization administrator, organization user, and the following five core entities. 

\begin{itemize}
    
    
    \item \textbf{LLM Services:} External large language model services (e.g., Microsoft's Copilot, OpenAI's ChatGPT, Perplexity) provide advanced AI capabilities for generating responses to user prompts.
    
    \item \textbf{Domain-specific LLM for Security (DLMS):} A customized LLM designed specifically for security purposes, handling sensitive tasks like confidentiality detection and compliance monitoring with organizational standards.
    
    \item \textbf{Sensitive Data Anonymizer (SDA):} An anonymous module that applies FPE algorithm to protect sensitive data detected by DLMS. This ensures that sensitive information is securely transformed into an encrypted format that preserves its structure, enabling safe transmission to the LLM without exposing raw sensitive data.
    
    \item \textbf{Model Monitoring Module:} A module for organization administrators to configure and manage the privacy protection system, including setting detection parameters, policies and monitoring system activity.

    \item \textbf{Database:} The database stores detected information, logs, and other essential data for the privacy protection system. The data is available for subsequent analysis, ensuring efficient retrieval and management.
\end{itemize}

For simplicity, Fig. \ref{fig:Framework} omits the Proxy and Integrator modules. The Proxy module serves as an intermediary layer intercepting and forwarding between users and external LLM services. The Integrator module manages data flow across components, processes information, and stores it in a database for monitoring and analysis. It ensures seamless integration, secures requests and responses, and enforces privacy policies.

The model training process mainly involves the following two modules.

\begin{itemize}
\item \textbf{Training Data Generation Module:}  The module processes sensitive documents by segmenting, annotating, and categorizing data using a predefined taxonomy and LLMs, creating structured training datasets and test sets for fine-tuning and evaluating DLMS.

\item \textbf{Model Training Module:} The module fine-tunes a base model using structured datasets and user-defined configurations, optimizing tasks like safety classification, category classification, and sensitive data extraction for creating DLMS. It employs frameworks like \textit{Llama Cookbook} and \textit{verl}, outputs model checkpoints, and monitors training dynamics using tools like \textit{Weights \& Biases}.
\end{itemize}

\subsection{Security Assumption}
The following defines the assumptions of trust and potential threats within the system.

\begin{itemize}
    \item Organization users are presumed to follow security protocols, but may inadvertently or deliberately expose sensitive data during interactions. 

    \item Public LLM services are assumed as {\it honest-but-curious}, meaning they adhere to agreed-upon protocols but may infer or retain sensitive information from user inputs.

    \item The integrator, DLMS, and SDA operate within the organization's secure boundary and are considered {\it trusted}. 

    \item It is assumed that the organization securely installs a Certificate Authority (CA) certificate when utilizing proxy tools, such as man-in-the-middle (MITM) proxies, to prevent unauthorized interception or modification of communications.

    \item The FPE algorithm (cf. \S \ref{Motivation for Format-Preserving Encryption}) is considered {\it secure} against known cryptanalytic attacks, following National Institute of Standards and Technology (NIST) recommendations \citep{FF3-1}. 
\end{itemize}

\begin{figure}
    \centering
    \includegraphics[width=\linewidth]{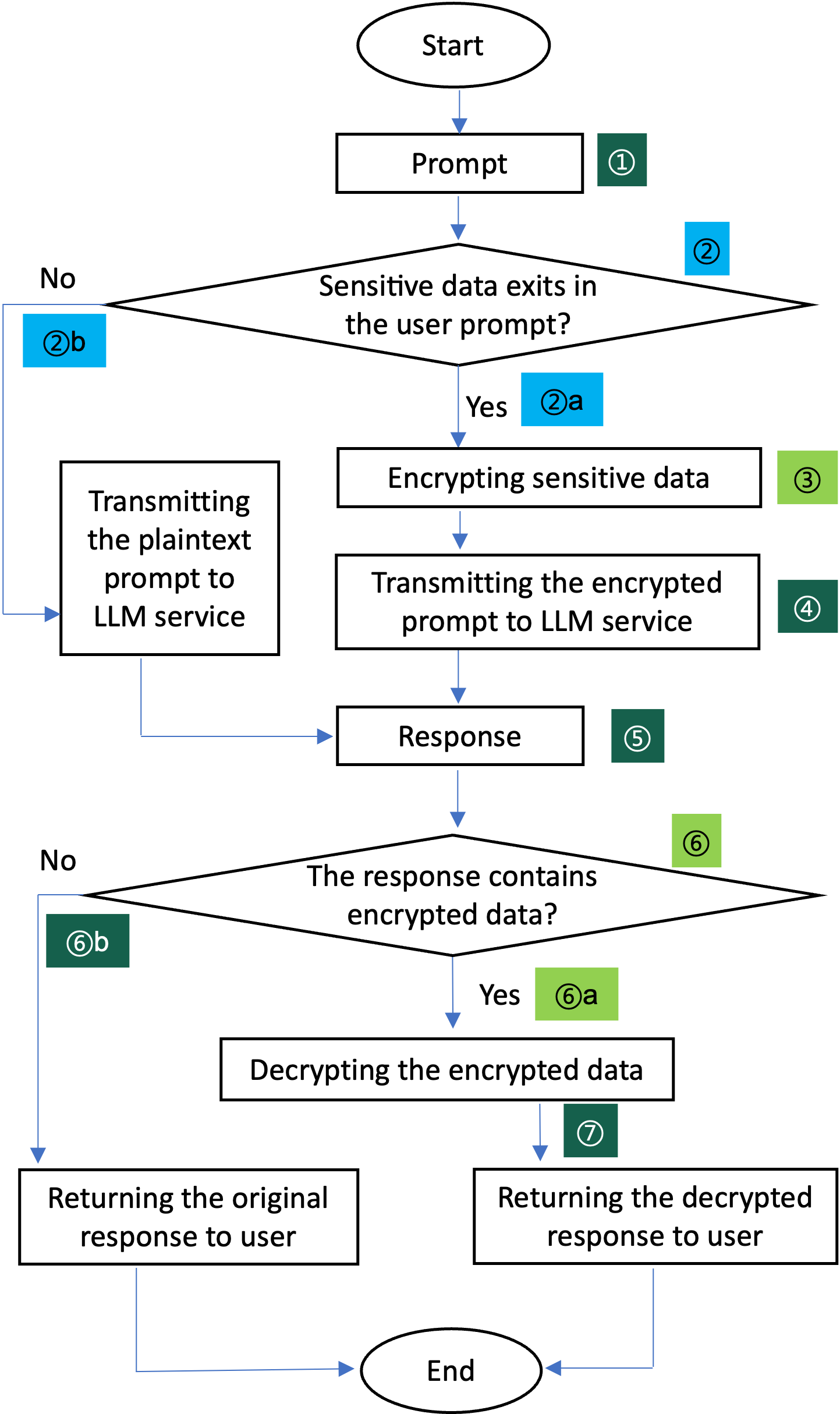}
    \caption{Simplified Workflow of LLM Access Shield. 
    }
    \label{fig:Workflow}
\end{figure}

\subsection{Workflow}
Fig. \ref{fig:Workflow} presents the workflow of secure communication with LLM services. 
The process involves analyzing the input prompt, encrypting sensitive data if necessary, processing the request via LLM services, and finally analyzing the response to determine the appropriate action. 
This workflow consists of seven key steps that ensure the safe handling of sensitive information. 

\begin{itemize}
    \item[] \textbf{\scalebox{0.85}{\Circled{1}} Prompt Submission:} 
    The organization user inputs a prompt.
    
    \item[] \textbf{\scalebox{0.85}{\Circled{2}} Prompt Analysis by DLMS:} 
    The DLMS analyzes the prompt to determine whether it contains sensitive data based on the organization's security policies. This may include personally identifiable information (PII), financial details, or confidential organizational data. 
    
    \begin{itemize}
        \item \textbf{{\scalebox{0.85}{\Circled{2a}}} Unsafe Prompt:}
        If the prompt is classified as ``unsafe", meaning it contains sensitive data, the DLMS identifies the categories of sensitive data detected (e.g., email address, phone number, bank account number).
        
        \item \textbf{{\scalebox{0.85}{\Circled{2b}}} Safe Prompt:}
        If the prompt is classified as ``safe", it is sent directly to the LLM services without modification.
    \end{itemize}
    
    \item[] \textbf{\scalebox{0.85}{\Circled{3}} Sensitive Data Encryption:} 
    In the case of {\scalebox{0.85}{\Circled{2a}}}, the SDA module encrypts the detected sensitive data using the FPE algorithm. 
    For instance, as shown in Fig. \ref{fig:Framework}, the phone number ``+7 323 406 7011" detected by DLMS is anonymized by the SDA module and transformed into ``+8 304 178 0394". 

    \item[] \textbf{\scalebox{0.85}{\Circled{4}} Processing by LLM Services:} 
    The encrypted or original prompt is forwarded to the LLM services.

    \item[] \textbf{\scalebox{0.85}{\Circled{5}} Response Generation:}
    The LLM services generate a response based on the provided prompt. 

    \item[] \textbf{\scalebox{0.85}{\Circled{6}} Sensitive Data Decryption:} 
    The SDA module verifies whether the response contains encrypted sensitive data:
    \begin{itemize}
        \item \scalebox{0.85}{\Circled{6a}}. If the response contains encrypted sensitive entities,  
        the SDA module decrypts them using the FPE decryption algorithm. A decrypted response is returned to the user in step \scalebox{0.85}{\Circled{7}}.
        
        \item \scalebox{0.85}{\Circled{6b}}. If the response does not contain encrypted data, the original response is returned to the user.
    \end{itemize}
\end{itemize}

\section{Experiment Setup}

\subsection{Taxonomy}

Different industries and organizations often define sensitive data entities using distinct taxonomies based on their specific privacy and security requirements. Our framework is designed to accommodate a wide range of customized taxonomies, providing flexibility to address diverse organizational needs.

For experimentation, we adopt a set of six commonly used sensitive data categories relevant to typical business-oriented companies: 
T1 \label{T1} (email address), T2 (personal identification (ID) number), T3 (phone number), T4 (fax number), T5 (bank account number), and T6 (monetary values). 
These categories are frequently encountered in corporate environments and are typically protected under organizational data protection policies. The sensitive data taxonomy and generation guidelines are provided in Appendix~\ref{appx:testing_data_generation}.

\subsection{Training and Testing Dataset}
We construct a structured input-output dataset containing both positive and negative samples. As shown in Table~\ref{tab:training_data_example}, each SFT training sample includes (1) a user message that probably contains sensitive data, (2) a ground-truth label indicating the safety of the message, (3) violated category codes (prefixed with ``T'') that represent the ground-truth set of the violated categories, and (4) and an explanation listing the ground-truth sensitive entities, separated by semicolons. We use the term ``Explanation'' as adopted in instruction-tuned datasets like LLaMA-Instruct.


\begin{table}[h!]
\centering
\caption{SFT training sample example.}
\label{tab:training_data_example}
\footnotesize
\resizebox{\columnwidth}{!}{%
\begin{tabularx}{\columnwidth}{@{}lX@{}}
    \toprule
    \textbf{Field} & \textbf{Example Value} \\
    \midrule
    Message & \texttt{Summarize this contract: Contract with Company A ... Fund value \$150,000 ... contact the customer at customer@gmail.com} \\
    Label & \texttt{unsafe} \\
    Violated Category Codes & \texttt{[T1, T6]} \\
    Explanation & \texttt{customer@gmail.com; 150,000} \\
    \bottomrule
\end{tabularx}%
}
\end{table} 

Table~\ref{tab:data-summary} summarizes the dataset, split into 60\% training (2,311 samples) and 40\% testing (1,542 samples).


\begin{table}[h!]
\centering
\caption{Summary of the training and testing datasets.}
\label{tab:data-summary}
\scriptsize
\resizebox{0.48\textwidth}{!}{
\begin{tabular}{l r r r r}
    \toprule
    \textbf{Metric} & \multicolumn{2}{c}{\textbf{Training Dataset}} & \multicolumn{2}{c}{\textbf{Testing Dataset}} \\
    \cmidrule(lr){2-3} \cmidrule(lr){4-5}
    & \textbf{Count} & \textbf{\%} & \textbf{Count} & \textbf{\%} \\
    \midrule
    Total msgs & 2,311 & 100.00\% & 1,542 & 100.00\% \\
    Safe msgs & 670 & 28.99\% & 452 & 29.31\% \\
    Unsafe msgs & 1,641 & 71.01\% & 1,090 & 70.69\% \\
    Single-labeled unsafe msgs & 1,345 & 58.20\% & 874 & 56.68\% \\
    Multi-labeled unsafe msgs & 296 & 12.81\% & 216 & 14.01\% \\
    \midrule
    \textbf{Category Breakdown} & & & & \\
    No Category: Safe & 670 & 28.99\% & 452 & 29.31\% \\
    T1: Email Address & 567 & 24.53\% & 384 & 24.90\% \\
    T2: Personal ID Number & 244 & 10.56\% & 161 & 10.44\% \\
    T3: Phone Number & 255 & 11.03\% & 179 & 11.61\% \\
    T4: Fax Number & 231 & 10.00\% & 133 & 8.63\% \\
    T5: Bank Account Number & 248 & 10.73\% & 174 & 11.28\% \\
    T6: Monetary Value & 410 & 17.74\% & 289 & 18.74\% \\
    \bottomrule
\end{tabular}
}
\end{table}

\subsection{Base Model and Training Methodology}
The framework supports any instruction-following language model as a base, allowing flexibility across downstream tasks. In this work, we fine-tune Llama-3.2-3B-Instruct to build DLMS, chosen for its lightweight architecture that balances efficiency and performance. In our experiments, the 3B model performs comparably to an 8B model, especially on small to medium-sized datasets, making it suitable for resource-constrained deployment.

We adopt two baseline training methods: (1) an SFT approach inspired by Llama-Guard \citep{inan2023llama}, and (2) an RFT method similar to DeepSeek-R1-Zero \citep{guo2025deepseek}, which uses rule-based rewards without SFT.

\section{Supervised Fine-Tuning (SFT)}
\label{sect:sft}


\subsection{Training Framework}

We use the \textit{Llama Recipes} repository (now renamed \textit{Llama Cookbook}) for SFT fine-tuning and utilize its formatting tools. Following Llama Guard \citep{inan2023llama}, we adopt a similar instruction and prompt format (see Appendix~\ref{appx:prompt_instructions}), but replace the original content safety taxonomy with our own sensitive data taxonomy and guidelines. Unlike Llama Guard, we do not apply the data augmentation strategy described in their work.

\subsection{Prompt Instruction}

We adopt a structured prompt format similar to Llama-Guard \citep{inan2023llama}, adapted to our sensitive data taxonomy and guidelines. Each prompt includes task instructions, a list of unsafe content categories, and a conversation segment, followed by a structured output. The output explicitly indicates whether the message is ``safe'' or ``unsafe'', lists the violated category codes, and extracts relevant unsafe entities. Detailed instruction templates for SFT are provided in Appendix~\ref{appx:prompt_instructions}.

\subsection{Training Configurations}


We train the DLMS-SFT-3B model by fine-tuning the pre-trained Llama-3.2-3B-Instruct \citep{inan2023llama} on our dataset of 2,311 samples, using our sensitive data taxonomy. Full-parameter fine-tuning is applied.

Training is conducted using a context length of $4,096$, over $3$ epochs, with an initial learning rate of $1\text{e}{-5}$ and a weight decay of $0.0$. We use a per-device batch size of $1$ with gradient accumulation steps of $2$, effectively simulating a larger batch size. The batching strategy is set to padding, ensuring uniform token lengths across batches. Mixed precision training is enabled, and we perform gradient clipping with a threshold of $1.0$ to stabilize training and prevent gradient explosion. Additionally, we employ an exponential learning rate decay with a factor of $\gamma = 0.85$ at the end of each epoch to improve optimization stability.

Validation is performed during training with a batch size of $1$, as specified by the \texttt{val\_batch\_size} parameter in the training script, and is enabled through the \texttt{run\_validation} flag. To improve memory efficiency, Fully Sharded Data Parallel (FSDP) is employed, with low-CPU mode enabled to minimize memory usage during checkpointing. The training process leverages fast kernels such as Flash Attention and Xformer for accelerated computation. The model is saved after fine-tuning for downstream use.

\section{Reinforcement Fine-Tuning (RFT)}
\label{sect:rft}

Although the Llama-Guard-style SFT approach demonstrates impressive performance on privacy policy taxonomy compliance, it suffers from limited generalizability and scalability, as it can only capture privacy taxonomy patterns present in the training data. In contrast, we aim to develop a more lightweight, flexible, and generalizable framework for policy compliance. 
Thus, we propose an \textit{analyze-then-decide} inference framework that integrates structured reasoning into the inference process for privacy policy compliance. Our approach uses reinforcement learning (RL), without SFT, to post-train base models for reasoning capabilities.


\subsection{Training Framework}

For the RFT approach, \texttt{verl} library, a state-of-the-art RL training framework optimized for LLMs, is utilized. \texttt{verl} provides flexible support for implementing reinforcement learning algorithms such as PPO, GRPO, and REINFORCE++, and integrates seamlessly with popular LLM infrastructures, including Hugging Face Transformers, vLLM, and DeepSpeed.

\subsection{Prompt Instruction}

To support the RFT process, we extend the SFT dataset with structured prompt instructions and reward model signals. This augmentation enables the model to learn to reason about privacy risks and to output decisions that align with privacy policy compliance objectives.

The prompt generation process employs either zero-shot or few-shot reasoning templates. Zero-shot templates provide task instructions without examples, whereas few-shot templates include several annotated examples to guide the model. Regardless of the template type, each prompt instructs the model to analyze the user's input for privacy violations and explain its reasoning. The reasoning process is encapsulated within \texttt{<analyze>...</analyze>} tags, while the final assessment, including safety status, violated categories, and sensitive entities (if applicable), is encapsulated within \texttt{<answer>...</answer>} tags. This structured format encourages the model to elaborate on its reasoning process, which is critical for RL optimization.

The detailed prompt instructions and examples used in the few-shot configuration are provided in Appendix~\ref{appx:prompt_instructions}, with the zero-shot configuration differing only by the exclusion of these examples. The output examples are shown in Appendix~\ref{appx:sample_outputs}. The processed dataset is saved in a format compatible with \texttt{verl} framework. Each dataset entry consists of the generated prompt, the corresponding reward model, and auxiliary metadata, such as dataset splits and indices. Each entry includes the following fields:

\begin{table}[h!]
\centering
\caption{RFT training sample example.}
\label{tab:rft_training_data_example}
\footnotesize
\resizebox{\columnwidth}{!}{%
\begin{tabularx}{\columnwidth}{@{}lX@{}}
    \toprule
    \textbf{Field} & \textbf{Example Value} \\
    \midrule
    Prompt & \texttt{<|begin\_of\_text|><|start\_header\_id|>user
    <|end\_header\_id|> Task: Check if there is unsafe content in 'User' messages in conversations according to our safety policy with the below categories. ... <analyze> Analyzing process here... </analyze> <answer> unsafe T1, T6 customer@gmail.com; 150,000 </answer>} \\
    Ability & \texttt{privacy\_risk\_analysis} \\
    Reward Model & \texttt{\{"style": "rule", "ground\_truth": \{"safety": "unsafe", "categories": ["T1", "T6"], "entities": "customer@gmail.com; 150,000"\}\}} \\
    Extra Info & \texttt{\{"split": "train", "index": 42\}} \\
    \bottomrule
\end{tabularx}%
}
\end{table}

\subsection{REINFORCE++ RL Algorithm}

In this work, we adopt the REINFORCE++ algorithm \citep{hu2025reinforce++}, an enhanced variant of the foundational REINFORCE algorithm to train reasoning version of DLMS. REINFORCE++ integrates key optimization techniques from PPO while maintaining the simplicity of the original REINFORCE framework by eliminating the need for a critic network. This design enables improved training stability and computational efficiency, making REINFORCE++ particularly well-suited for large-scale applications.

\subsection{Reward Modelling}

To train DLMS under the RFT framework, we design a rule-based reward mechanism that evaluates the model's ability to identify privacy risks in user prompts. The reward model evaluates three complementary aspects of the model's output: format compliance, classification accuracy, and entity-level extraction accuracy. These components ensure that the model generates outputs that are both structurally coherent and semantically aligned with the ground truth. Below, we describe each reward component in details.

\textit{Format reward.} The format reward incentivizes the model to generate outputs adhering to a predefined structured format. Specifically, the reasoning steps must be enclosed within \texttt{<analyze>...</analyze>} tags, and the final prediction must be enclosed within \texttt{<answer>...</answer>} tags. These tags must appear exactly once, in the correct sequence, and without any extraneous content outside these tags. A reward of $+2$ is granted if the format is correct, and a penalty of $-2$ is applied otherwise. The format reward function is defined as:
\[
R_{\text{fmt}}(y) =
\begin{cases}
2, & \text{if the format matches,} \\
-2, & \text{if the format is incorrect,}
\end{cases}
\]
where $y$ denotes the model's output. The validation process ensures that the \texttt{<analyze>} tags precede the \texttt{<answer>} tags, the content between \texttt{<analyze>} and \texttt{<answer>} tags is uninterrupted, and any text following the closing \texttt{</answer>} tag is either empty or matches an expected end-of-sequence (EOS) token.

\textit{Safety status reward.} The safety status reward evaluates the model's ability to correctly classify user prompts as either ``safe'' or ``unsafe.'' A correct classification yields a reward of $+1$, while an incorrect classification or an invalid output (e.g., ``unknown'') incurs a penalty of $-1$. The safety status reward function is defined as:
\[
R_{\text{safety}}(y, y^*) =
\begin{cases} 
1, & \text{if } y_{\text{safety}} = y^*_{\text{safety}}, \\
-1, & \text{if } y_{\text{safety}} \neq y^*_{\text{safety}}, \text{ or } \\ 
   & y_{\text{safety}} = \text{``unknown'',}
\end{cases}
\]
where $y_{\text{safety}}$ represents the predicted safety status, and $y^*_{\text{safety}}$ is the ground truth safety label.

\textit{Category code reward.} For prompts classified as ``unsafe,'' the model is required to identify the specific category codes corresponding to the violated privacy-sensitive data types (e.g., ``T1'' for email address, ``T2'' for personal ID). The category code reward evaluates the alignment between the predicted and ground-truth category sets, assigning scores based on their degree of overlap. A perfect match yields a reward of $+2$. If the predicted set forms a proper, non-empty subset of the ground truth, a reward of $+1$ is granted. In all other cases, a penalty of $-1$ is applied. The reward function is formally defined as:
\[
R_{\text{cat}}(C_{\text{pred}}, C_{\text{gt}}) =
\begin{cases}
2, & \text{if } C_{\text{pred}} = C_{\text{gt}}, \\
1, & \text{if } C_{\text{pred}} \neq \emptyset \text{ and } C_{\text{pred}} \subset C_{\text{gt}}, \\
-1, & \text{otherwise,}
\end{cases}
\]

\noindent where $C_{\text{pred}}$ and $C_{\text{gt}}$ denote the predicted and ground-truth sets of category codes, respectively.

\textit{Entity extraction reward.} The entity-level extraction reward evaluates the model's precision in identifying specific strings of sensitive data entities (e.g., email addresses or personal IDs) within the user prompt. A perfect match between the predicted and ground truth entities yields a reward of $+4$. A partial match, where the predicted entities form a proper subset of the ground truth entities, earns a reward of $+2$. All other cases incur a penalty of $-1$. The entity-level reward function is defined as:
\[
R_{\text{ent}}(E_{\text{pred}}, E_{\text{gt}}) =
\begin{cases}
4, & \text{if } E_{\text{pred}} = E_{\text{gt}}, \\
2, & \text{if } E_{\text{pred}} \neq \emptyset \text{ and } E_{\text{pred}} \subset E_{\text{gt}}, \\
-1, & \text{otherwise,}
\end{cases}
\]
where $E_{\text{pred}}$ and $E_{\text{gt}}$ denote the sets of predicted and ground-truth sensitive entities, respectively.

\textit{Final reward aggregation.} The final reward score is computed as the sum of the individual reward components:
\[
R_{\text{total}}(y, y^*) = R_{\text{fmt}}(y) + R_{\text{safety}}(y, y^*) + R_{\text{cat}}(y, y^*) + R_{\text{ent}}(y, y^*).
\]

All four reward components are active during training. This aggregated reward encourages the model to produce outputs that are not only structurally well-formed, but also accurate in safety classification, category code identification, and sensitive entity extraction. By combining these metrics, the reward model provides a comprehensive evaluation framework for privacy risk detection in user prompts.

\subsection{Curriculum Learning (CL)}

Curriculum Learning (CL) is a training paradigm in which models are exposed to tasks of increasing complexity, thereby improving convergence behavior and overall performance. By structuring learning progressively, CL helps models converge to better local optima, especially when global optima are difficult to attain. Task difficulty in CL can be predefined or dynamically determined during training. Logic-RL \citep{xie2025logic} illustrates the efficacy of CL in RFT.

In this study, the task of sensitive data detection is naturally decomposed into a hierarchy of three levels: (1) safety status classification, a binary task to determine whether a user prompt is “safe” or “unsafe”; (2) category code classification, a multi-label task to identify categories violated by unsafe prompts; and (3) sensitive data extraction, a sequence-level task to extract specific sensitive entities such as Personally Identifiable Information (PII) and Business Identifiable Information (BII).

To address this task hierarchy, we propose a three-stage Curriculum Learning (CL) framework to incrementally fine-tune the DLMS model. This framework leverages the structured progression of task complexity to improve performance and mitigate catastrophic forgetting. Each stage targets a specific subtask:

\begin{enumerate}
    \item Safety status classification:
    The model is initially fine-tuned on binary safety classification, emphasizing structured output formats inspired by the “analyze-answer” framework in DeepSeek-R1-Zero. This stage establishes a foundation for reasoning and output consistency.

    \item Category code classification:
    Building on Stage 1, the model is fine-tuned for multi-label classification to identify ``unsafe" categories, refining its understanding of safety while learning category predictions.

    \item Sensitive data extraction:
    In the final stage, the model performs sequence-level predictions to extract sensitive entities (e.g., PII, BII) while retaining proficiency in the prior tasks.
\end{enumerate}

The model is fine-tuned for one epoch per stage, resulting in a lightweight training process of three epochs. This progressive design aligns with the principle of scaffolding, where earlier tasks provide a foundation for mastering more complex ones, ensuring computational efficiency while achieving robust performance.

In the context of training DLMS for sensitive data detection via CL, we adopt a stage-specific reward function design. This approach ensures that the model is incentivized to focus on the relevant sub-task at each stage, while gradually progressing towards the final goal. The reward system follows the same principles as those used in the SFT setting; yet it is selectively activated and weighted across stages to reflect task complexity. 
The only notable difference lies in the \textit{format reward}, where the penalty for incorrect output formatting evolves over stages: no penalty is applied at Stage~1 (reward = 0), while a stronger penalty of $-2$ is assigned at Stages~2 and~3 to enforce stricter adherence to format as tasks become more complex.

In summary, Stage~1 rewards focus on format and safety classification; Stage~2 introduces category code rewards; and Stage~3 adds entity extraction rewards. This stage-specific reward scheduling ensures that the model is gradually guided to master increasingly complex tasks, in line with the principles of curriculum learning.

\subsection{Training Configurations}



The model is trained with a learning rate of $1\text{e-}6$ for 3 epochs. Following the default settings of the \texttt{verl} framework, the learning rate warm-up ratio is set to 0. The total batch size is 4 for training and 16 for validation. Each training batch is further divided into mini-batches of size 8, with a micro-batch size of 4 per GPU. The maximum sequence length is set to 4,096 tokens for both prompts and responses, enabling the model to process long-form reasoning tasks effectively.

To optimize GPU memory usage, both gradient checkpointing and padding removal are enabled, with the latter improving efficiency during token-level processing. For the actor model, FSDP offloading of parameters and optimizer states is disabled to prioritize computational speed. In contrast, parameter offloading is enabled for the reference model to reduce memory consumption. The FSDP wrap policy follows the default configuration provided by Hugging Face.

To generate rollouts efficiently, the vLLM hybrid engine is used with a tensor parallel size of 4 and a GPU memory utilization ratio of 0.4 for key-value cache storage. Rollout sampling is performed with a temperature of 0.7, and the same temperature is applied during validation for consistency. The sampling strategy uses the default configurations for \texttt{top\_k} and \texttt{top\_p}, which are set to $-1$ (no top-k filtering) and $1.0$ (no truncation of probability mass), respectively. These settings ensure that the full token probability distribution is considered during sampling.

The KL divergence regularization coefficient $\beta$ is set to 0.001, and a low-variance KL divergence loss is used to stabilize policy updates. The model generates 8 rollouts per prompt, enabling exploration of diverse responses. For log probability computations, a micro-batch size of 8 is used for both the rollout and reference models.

Reward estimation follows a purely rule-based approach, and no separate reward model is employed, thereby simplifying the training pipeline.  Generalized Advantage Estimation (GAE) is used for advantage computation with hyperparameters $\lambda = 1$ and $\gamma = 1$, which are the default values in the \texttt{verl} framework. These parameters ensure that the advantage estimation fully weights long-term rewards without discounting or bias. The critic warm-up phase is disabled, and training begins directly with policy updates.

Model checkpoints are saved every 100 training steps, and validation is performed at the same interval. Although the \texttt{verl} framework defaults to validating every 2 steps, this frequency is increased to reduce computational overhead. The model checkpoint achieving the highest validation reward is selected for evaluation. In the curriculum learning training scheme, the model checkpoint of the previous stage achieving the highest validation reward is selected for the training of the next stage.

\section{DLMS Evaluation Setup}
\label{sect:evaluation_setup}

\subsection{Model Versions}

Three model variants, all of which use parameter fine-tuning without parameter-efficient fine-tuning (PEFT), are evaluated.
All models are fine-tuned using the same sensitive data taxonomy, with Llama-3.2-3B-Instruct as the base model, and trained on a dataset comprising 2,311 samples.

\begin{enumerate}
    \item \textbf{DLMS-SFT}: The base model is fine-tuned with three epochs using  SFT only.
    \item \textbf{DLMS-RFT}: The base model is fine-tuned with three epochs using RFT only.
    \item \textbf{DLMS-RFT-CL}: The base model is fine-tuned using a three-stage CL strategy, where each stage is trained for one epoch.
\end{enumerate}

\subsection{Evaluation Metrics}
We evaluate model performance across three levels: safety, category, and entity.

At the \textbf{safety level}, which is formulated as a binary classification task, we use standard metrics including Accuracy, F1-Score, and Area Under the Precision-Recall Curve (AUPRC) to assess the model’s ability to distinguish between ``safe'' and ``unsafe'' prompts.

At the \textbf{category level}, since a single prompt may involve multiple sensitive categories, we treat the task as multi-label classification and report Subset Accuracy, Hamming Accuracy, and Multi-Label F1-Score.
Subset Accuracy is the most stringent metric used, measuring the proportion of instances where the predicted label set exactly matches the ground truth. It is formally defined as $\frac{1}{N} \sum_{i=1}^{N} I(y_i = \hat{y}_i)$, where \(N\) represents the number of instances, \(y_i\) and \(\hat{y}_i\) denote the true and predicted sets of labels for instance \(i\), respectively, and \(I\) is an indicator function that returns 1 if \(y_i = \hat{y}_i\) and 0 otherwise. In contrast, Hamming Accuracy provides a more lenient metric by evaluating prediction correctness independently for each label. It is defined as $\frac{1}{N \times L} \sum_{i=1}^{N} \sum_{l=1}^{L} I(y_{il} = \hat{y}_{il})$, where \(L\) is the total number of labels, and \(y_{il}\) and \(\hat{y}_{il}\) represent the true and predicted values for label \(l\) of instance \(i\), respectively

At the \textbf{entity level}, we use the \textit{Privacy Hiding Rate} (PHR)~\citep{24-SXHTHZ-ProSan} to quantify the effectiveness of sensitive entity obfuscation. PHR is defined as the proportion of sensitive entities that are successfully anonymized by the model: $\frac{N_{\text{hidden}}}{N_{\text{total}}}$, where \(N_{\text{total}}\) is the total number of sensitive entities identified in the prompts, and \(N_{\text{hidden}}\) is the number of those entities that are successfully anonymized by the model.

\section{DLMS Evaluation Results}
\label{sect:evaluation}

\begin{table*}[t]
    \centering
    \caption{Privacy protection performance.}
    \label{tab:performance-metrics-privacy}
    \scriptsize
    \begin{tabularx}{\textwidth}{@{}l|XXX|XXXX|X@{}}
        \toprule
        \textbf{Model} & \multicolumn{3}{c|}{\textbf{Safety Level}} & \multicolumn{4}{c|}{\textbf{Category Level}} & \textbf{Entity Level} \\
        \cmidrule(lr){2-4} \cmidrule(lr){5-8} \cmidrule(lr){9-9}
        & \textbf{Accuracy} & \textbf{F1-Score} & \textbf{AUPRC} 
        & \textbf{Hamming Accuracy} & \textbf{Subset Accuracy} & \textbf{Multi-Label F1-Score} & \textbf{AUPRC} 
        & \textbf{Privacy Hiding Rate} \\
        \midrule
        DLMS-SFT               & \textbf{0.935} & \textbf{0.918} & \textbf{0.962} & 0.922 & 0.696 & 0.707 & 0.733 & \textbf{0.839} \\
        DLMS-RFT                & 0.910 & 0.882 & 0.946 & 0.919 & 0.660 & 0.701 & 0.697 & 0.555 \\
        DLMS-RFT-CL             & 0.916 & 0.891 & 0.950 & \textbf{0.928} & \textbf{0.698} & \textbf{0.735} & \textbf{0.742} & 0.533 \\
        Llama-3.2-3B-Instruct   & 0.468 & 0.462 & 0.869 & 0.832 & 0.303 & 0.268 & 0.336 & 0.064 \\
        \bottomrule
    \end{tabularx}
\end{table*}

\begin{table*}[t]
    \centering
    \caption{Privacy protection performance of different model checkpoints from curriculum learning.}
    \label{tab:performance-metrics-privacy-stages}
    \scriptsize
    \begin{tabularx}{\textwidth}{@{}l|XXX|XXXX|X@{}}
        \toprule
        \textbf{Model} & \multicolumn{3}{c|}{\textbf{Safety Level}} & \multicolumn{4}{c|}{\textbf{Category Level}} & \textbf{Entity Level} \\
        \cmidrule(lr){2-4} \cmidrule(lr){5-8} \cmidrule(lr){9-9}
        & \textbf{Accuracy} & \textbf{F1-Score} & \textbf{AUPRC} 
        & \textbf{Hamming Accuracy} & \textbf{Subset Accuracy} & \textbf{Multi-Label F1-Score} & \textbf{AUPRC} 
        & \textbf{Privacy Hiding Rate} \\
        \midrule
        DLMS-RFT-CL (stage 1 only)   
            & \textbf{0.965} & \textbf{0.958} & \textbf{0.984} 
            & 0.864 & 0.582 & 0.531 & 0.521 
            & \textbf{0.545} \\
        DLMS-RFT-CL (stages 1 \& 2) 
            & 0.951 & 0.941 & 0.976 
            & 0.922 & \textbf{0.712} & 0.703 & 0.660 
            & 0.493 \\
        DLMS-RFT-CL                          
            & 0.916 & 0.891 & 0.950 
            & \textbf{0.928} & 0.698 & \textbf{0.735} & \textbf{0.742} 
            & 0.533 \\
        \bottomrule
    \end{tabularx}
\end{table*}

\subsection{Performance on Privacy Risk Detection}

We use a zero-shot prompting method during model evaluation on the privacy risk detection task. This involves category names and their corresponding descriptions into the prompt at inference time.
Table~\ref{tab:performance-metrics-privacy} summarizes the performance of three DLMS models and the base model (Llama-3.2-3B-Instruct). For each metric, the highest-performing value is highlighted.

At the \textbf{safety level}, all DLMS models significantly outperform the baseline Llama-Guard-3-8B across Accuracy, F1-Score, and AUPRC, confirming their strong ability to detect whether a prompt poses a privacy risk. Among them, DLMS-SFT achieves the highest scores, slightly outperforming DLMS-RFT and DLMS-RFT-CL. Interestingly, DLMS-RFT-CL performs almost identically to DLMS-RFT, suggesting that CL has limited impact at this level.

At the \textbf{category level}, all DLMS models again outperform the baseline, demonstrating strong multi-label classification capabilities. DLMS-RFT-CL achieves the best performance across all metrics, indicating that CL may provide additional benefits for category-level generalization. DLMS-SFT and DLMS-RFT perform comparably on this task.

At the \textbf{entity level}, DLMS-SFT achieves the best Privacy Hiding Rate ($0.839$), significantly outperforming both DLMS-RFT ($0.555$) and DLMS-RFT-CL ($0.533$). In contract, the base model achieves only $0.064$ under the same zero-shot prompting setup. This highlights DLMS’s effectiveness in privacy-preserving inference.

Although the RFT-based models underperform SFT in this task, this outcome is partially due to the conservative reward design used during training. To reduce overprediction behaviors observed in early RFT runs, we introduced penalties for both category-level and entity-level overprediction. While this strategy successfully mitigated reward hacking, it also led the model to become overly cautious, resulting in lower recall and reduced Privacy Hiding Rates. We hypothesize that these conservative behaviors stem from an imbalanced reward design that discourages borderline predictions. A more fine-grained reward function that better balances false positives and false negatives could potentially improve the performance of RFT models on entity-level tasks.

\subsection{Evaluating the Impact of CL}

Table~\ref{tab:performance-metrics-privacy} suggests that the CL training strategy leads to mild improvements in safety status classification and category code extraction tasks. To further examine the impact of CL, we analyze the performance of three intermediate RFT model checkpoints from different CL stages, as reported in Table~\ref{tab:performance-metrics-privacy-stages}: (1) the best-performing model on the validation set after Stage 1 training (Stage-1 model), (2) the best checkpoint after Stages 1 and 2 (Stage-2 model), and (3) the final model trained through all three stages (Stage-3 model, i.e., DLMS-RFT-CL). For each metric, the best-performing value across all checkpoints is highlighted.

``Catastrophic forgetting'' is observed when the RFT model is trained sequentially by CL on multi-level privacy detection. The stage 2 and stage 3 models exhibit degraded performance on safety status classification, despite being trained on this task throughout all stages. This suggests that later-stage training may interfere with previously learned capabilities.

\section{Adaptability to Non-Taxonomy Policies}
\label{sect:adaptability}

The adaptability of the framework to non-taxonomy and general-purpose privacy policies is evaluated. By leveraging few-shot prompting, the DLMS enables dynamic and real-time policy enforcement, ensuring compliance with organizational privacy guidelines even in the absence of explicit training on specific policies.

To the best of our knowledge, there are no publicly available benchmarks designed to assess LLM's capability of privacy policy compliance. To address this gap, we curated a set of four authentic privacy policies extracted from the security policy documentation of an anonymous organization's Data Protection Office. These policies were selected to reflect real-world scenarios involving sensitive information. An example policy for a binary classification task is like: {\it ``Secret information is information of sensitive nature or having strategic values. ... Examples include passwords and cryptographic keys. The following usages are not permitted: Disseminating sensitive or confidential information of Company''.} Again, we generated a testing dataset of non-compliant user prompts containing unsafe content with the assistance of uncensored LLMs. See Table~\ref{tab:summary-policy-datasets-user-query} in Appendix~\ref{appx:testing_data_generation} for the summary statistics of the testing dataset for policy compliance of user prompts, and Table~\ref{tab:privacy_policies} in Appendix~\ref{appx:taxonomy_descriptions} for policy definitions of the four testing policies.

For evaluation metrics, the model's performance using the same groups of metrics at the safety level is assessed. The results of the evaluation are summarized in Table~\ref{tab:performance-metrics-privacy-policies}. The evaluation metric values in the table are the weighted average across all four testing policies.

\begin{table}[h!]
    \centering
    \caption{Adaptability to non‐taxonomy privacy policies via prompting.}
    \label{tab:performance-metrics-privacy-policies}
    \scriptsize
    \begin{tabularx}{\columnwidth}{@{}l X X X@{}}
        \toprule
        \textbf{Model} & \textbf{Accuracy} & \textbf{F1 Score} & \textbf{AUPRC} \\
        \midrule
        DLMS-SFT (1 epoch, FPFT)                     & 0.564 & 0.455 & 0.769 \\
        DLMS-SFT (1 epoch, LoRA)                     & 0.758 & 0.749 & 0.832 \\
        DLMS-RFT-CL (stage 1 only)          & \textbf{0.862} & \textbf{0.861} & \textbf{0.898} \\
        Llama-3.2-3B-Instruct                        & 0.713 & 0.706 & 0.801 \\
        \bottomrule
    \end{tabularx}
\end{table}

As we only assess the model's generalization capability to the first level of the privacy risk identification task, i.e., identify the safety status of the given prompt, we compare the model checkpoint of highest validation reward after stage 1 of CL training only (stage 1 model) against the DLMS-SFT models that were fine-tuned for 1 epoch using and not using Low-Rank Adaptation (LoRA) \citep{hu2022lora}. We find that the stage 1 model has the best performance among the models, which demonstrates the potential to improve generalization capability of RFT approach and the test-time compute scaling of ``analyze-then-decide'' paradigm.

\section{Utility-Preservation Encryption}

To enable secure and effective downstream processing following sensitive data detection by DLMS, we adopt a token-level encryption-based anonymization strategy. Specifically, instead of encrypting the entire prompt, we selectively encrypt sensitive entities detected within unsafe user inputs. This approach ensures targeted protection while preserving the overall utility of the prompt. 
Privacy-preserving inference in LLMs presents a fundamental trade-off between minimizing privacy leakage and preserving input utility. Zhang et al.~\citep{zhang2025no} formalize this dilemma as the \textit{No Free Lunch Theorem} for privacy-preserving inference, which states that stronger privacy guarantees often lead to reduced utility in model outputs.

In this work, we define \textbf{utility} as the preservation of the \textit{semantic context} of user prompts, i.e., the meaning conveyed by words and phrases rather than their surface forms~\citep{DBLP:conf/emnlp/ChuS0024}. This is essential for LLMs, which rely on semantic coherence to produce accurate and contextually appropriate outputs. Therefore, anonymization should protect sensitive content without significantly altering the input’s meaning.

\subsection{Motivation for Format-Preserving Encryption}
\label{Motivation for Format-Preserving Encryption}
Traditional encryption algorithms such as AES~\citep{AES} produce ciphertexts that completely obscure the structure of the original input, replacing sensitive tokens with opaque, unstructured strings. While this ensures strong confidentiality, it disrupts the syntactic and structural integrity of the prompt, which can lead to downstream processing failures or degraded LLM performance.

To address these limitations, we adopt Format-Preserving Encryption (FPE) as a utility-preserving anonymization technique. FPE produces ciphertexts that maintain the same format—length, character classes, and delimiters—as the original input~\citep{09-BRR+FPE}. 
As illustrated in Table~\ref{tab:fpe-vs-aes}, FPE preserves token-level formatting across all entity types, whereas AES produces opaque and inconsistent ciphertexts. This format retention is vital for schema validation and downstream interpretability, making FPE particularly suitable for LLM-based applications that rely on textual coherence and structural fidelity.

The NIST specifies two standard algorithms for FPE: FF1 and FF3-1~\citep{FF3-1}. FF1 offers greater flexibility, supporting a wider range of input lengths and tweak sizes. In contrast, FF3-1 reduces the number of encryption rounds from ten to eight, achieving higher throughput. In our experiments, we adopt FF3-1 for its performance benefits. The FPE encryption and decryption processes are defined as follows: let $K$ denote the encryption key, $T$ the tweak, $X$ the plaintext, and $Y$ the ciphertext. The tweak $T$ does not need to be secret, but using variable tweaks across encryptions enhances security.  The ciphertext $Y$ retains the same length as the plaintext $X$. The encryption function is denoted as $\text{FPE.Encrypt}(K, T, X)$, and the decryption function as $\text{FPE.Decrypt}(K, T, Y)$. For a fixed key and tweak, decryption is the inverse of encryption, satisfying $\text{FPE.Decrypt}(K, T, \text{FPE.Encrypt}(K, T, X)) = X$.

\begin{table*}[ht]
\centering
\caption{Comparison of FPE and AES ciphertexts.}
\label{tab:fpe-vs-aes}
\renewcommand{\arraystretch}{1.2}
\scriptsize
\begin{tabular}{|l|p{2.2cm}|p{2.2cm}|p{6.5cm}|}
\hline
\textbf{No Category} & \textbf{Plaintext} & \textbf{FPE Ciphertext} & \textbf{AES Ciphertext} \\
\hline
T1: Email Address & \texttt{tinavang@support.org} & \texttt{UbuEzHYT@HAyfwmn.com} & \texttt{vzbblybt5iKs0AURAp9uKXKo4IDGEnS7QD3/enbhUVlqNSHkDGhezfA6kzeX2apt} \\
\hline
T2: Personal ID Number & \texttt{B 987 654 3} & \texttt{I 1aA aL7 d} & \texttt{QqljoR0E6sEzv5AqiWv70T2h2tjawPNK4aBPI8JZjKc=} \\
\hline
T3: Phone Number & \texttt{+86 13945093743} & \texttt{+92 43651064790} & \texttt{pn7ZmoVELmMhINcy1ePJHGxvWY3Wrw41boMh+wxf7f0=} \\
\hline
T4: Fax Number & \texttt{(853) 3406-2802} & \texttt{(645) 2766-3262} & \texttt{McJ3oiDASSialUbySpLyxNmOTA1etBGrLmJje74NBgE=} \\
\hline
T5: Bank Account Number & \texttt{DE89370400440532013000} & \texttt{DE79420195675275016155} & \texttt{IWs6jpT+sFnO7u1LCKnUnsw3MalkJ92qw60n41QkS5dp/m0g5lXYTGLRKlHVycBz} \\
\hline
T6: Monetary Value & \texttt{1,452,500} & \texttt{6,423,095} & \texttt{zPLGuKCLJrVdkPkKVU5Ku119xPEo7RgaZ/6bQ8Mg43Y=} \\
\hline
\end{tabular}
\end{table*}

\subsection{Effectiveness of FPE}
Beyond structural preservation, FPE also demonstrates strong empirical performance in maintaining the semantic utility of prompts for downstream LLM tasks. Appendix Fig.~\ref{fig:Response_Comparison} shows that Copilot produces nearly identical responses to FPE-encrypted and plaintext prompts, indicating that entity-level encryption does not compromise model understanding or output quality.

\section{Discussions}
\label{sect:discussion}

\subsection{Rule-Based Reward Design for DLMS-RL for Policy Compliance}

Privacy regulations in industries like technology, finance, and healthcare are complex, requiring a deep understanding to design effective compliance systems. Research on automated privacy and security systems has shown that formalizing policies can detect violations, automate enforcement, and evaluate alternative designs. This is particularly critical in highly regulated domains governed by frameworks such as GDPR in Europe or HIPAA in the United States, where policies impose strict constraints on handling sensitive data, such as Electronic Health Information (EHI). The increasing complexity of these frameworks has driven demand for Regulatory Technology (RegTech) solutions that operationalize compliance at scale.

Our trials with RFT of DLMS indicate that fine-grained compliance relies on formalizing policies as logic-based rules. Expressing policies as logic programs enables automatic generation of reward modeling procedures for RFT, encoding interpretable rules to evaluate compliance across multiple levels of granularity. This approach supports dynamic reward functions that adapt to specific policy requirements, such as prioritizing sensitive data or balancing trade-offs between compliance metrics. While not exhaustive, our findings suggest that rule-based reward modeling facilitates customizable, resource-efficient enforcement of privacy policies, particularly in resource-constrained or domain-specific settings.
Below are concrete examples that illustrate the utility of rule-based reward modeling in real-world scenarios:

\textit{Differentiated safety priorities.} Consider a social media company that prioritizes blocking unsafe prompts to prevent misuse of its platform, such as the generation of harmful or offensive content. In contrast, a financial institution focuses on ensuring that sensitive information, such as customer account numbers or transaction details, is encrypted. For the social media company, the reward model would emphasize binary classification of safety, while for the financial institution, it would prioritize sensitive entity extraction.

\textit{Balancing false positives and false negatives.} A cybersecurity firm may aim to minimize false positives (e.g., unnecessary alerts that disrupt workflows), as these can overwhelm its monitoring systems and reduce operational efficiency. On the other hand, a healthcare organization might prioritize minimizing false negatives, as these represent undetected privacy risks that could lead to severe regulatory penalties or patient data breaches. Rule-based reward modeling allows for these preferences to be encoded directly into the training process, tailoring the model's behavior to the specific needs of the organization.


\textit{Optimizing evaluation metrics.} An AI research lab deploying a model for internal use may require the model to achieve an F1-score $>$ 0.9 to balance precision and recall for its compliance tasks. However, a legal tech firm might prioritize achieving the highest possible precision to ensure that no non-compliant actions are inadvertently flagged as compliant. The reward modeling system can be tuned to optimize for the metric most relevant to the organization’s objectives.

\subsection{Dynamic Policy Compliance via Prompting}

The overfitting of domain-specific models to specific taxonomies of “safe” and “unsafe” undermines their stability and generalization across scenarios, making them unsuitable for rapidly evolving privacy policies or high-risk applications. Fine-tuning models for real-time and dynamic policy compliance with ad hoc collection of training data is impractical due to the high cost, latency, and resource demands. Businesses require adaptable frameworks capable of handling policy dynamics without constant retraining.

In Section \ref{sect:adaptability}, we demonstrate that DLMS can address this issue and perform a non-taxonomy privacy policy classification task using a few-shot prompting method. This approach leverages DLMS’ generalization capabilities to dynamically interpret and enforce policies, reducing deployment costs and enabling real-time compliance. Unlike traditional fine-tuning methods, prompting provides a flexible and cost-effective solution for adhering to diverse and evolving privacy regulations, such as GDPR or HIPAA.

Dynamic and context-sensitive compliance is critical in domains, such as healthcare, finance, and technology, where regulations vary across jurisdictions and evolve rapidly. Traditional fine-tuned models often fail to generalize beyond narrowly defined taxonomies, leading to instability and inaccuracies in high-stakes applications. By contrast, a prompting-based methodology enables DLMS to adapt to unseen policies, supporting real-time compliance with minimal disruption.

\subsection{Transparency and Traceability}

The black-box nature of LLM inference poses a significant challenge in regulated domains such as privacy, security, and compliance, where transparency and traceability are critical. Regulations like the GDPR the HIPAA emphasize the need for explainability in automated decision-making systems, particularly when these systems are deployed in sensitive applications. Despite their high performance, LLMs often fail to meet these requirements due to the opaque nature of their inference processes, limiting their applicability in policy compliance and enforcement in critical business areas.

To address this limitation, we propose an analyze-then-decide inference framework that integrates structured reasoning into the inference process, enhancing both transparency and traceability. This framework is implemented through our DeepSeek R1-Zero-like RFT approach, which trains models to provide human-readable justifications for their outputs. Specifically, the model is designed to output its reasoning process enclosed within \verb|<analyze>| tags before delivering a final decision enclosed within \verb|<answer>| tags. By explicitly generating reasoning steps, the model allows regulators, auditors, and operators to trace how specific decisions were derived, thereby satisfying compliance requirements for explainability.

\section{Related Work}
\label{sect:related_work}

Protecting sensitive data during training, including both pre-training and fine-tuning, is a key strategy for ensuring privacy in LLMs. Homomorphic encryption (HE) \cite{16-RNKKM-Cryptonets-HE-ICML,22-CBH+THE-X-PPInferenceHE,22-HLC+Iron,23-LL-LLM-HE}, secure multi-party computation (MPC) \cite{23-HLL+Ciphergpt-MPC,23-AFAS-Privformer-MPC-EuroSP,23-DLZ+Puma,23-GJM+2PC-FSS,23-LHG+Bumblebee-HE}, trusted execution environments (TEEs) \cite{shumailov2025trusted,frikha2024obfuscatune,gim2024confidential}, and differential privacy (DP) \citep{22-RMJB-EWTune-DP,22-SSCZJY-JustFineTuneTwice-DP,22-WGX-DP,22-MDPSGZ-DP,23-DYCWHS-DP-Forward-ccs,23-MYHYP-Split-and-Denoise-DP,23-INS+PrivChatGPT-DP,23-TCQ+Privinfer,23-LTL-RAPT,24-HWZ+ICLR-DP,23-UHG+PrivChatGPT} are widely adopted techniques for achieving this goal, and have been comprehensively reviewed in several recent surveys \cite{23-LWSY+Survey,24-EW-Survey,24-YLX+LLMPrivacy-Survey,24-WZL+LLMPrivacy-Survey,24-DAW-LLM-Survey}. Therefore, in this work, we focus on introducing recent studies specifically aimed at preserving sensitive information during user–LLM interactions. 

PP-TS \cite{23-KQY+text-sanitization} first identifies and removes sensitive information from user inputs through multiple rounds of text sanitization. The sanitized input is then processed by a remote LLM, and the sensitive content is reintegrated into the model’s output using local plaintext–ciphertext mappings.
ConfusionPrompt \cite{24-MYYYP-ConfusionPrompt} and Instance-Obfuscated Inference (IOI) \cite{yao2024privacy} employ decomposition techniques, pseudo-prompts, and decision resolution mechanisms to obscure sensitive inputs while reconstructing the desired outputs. These methods provide practical solutions but must carefully navigate trade-offs between privacy guarantees and prompt usability.
HaS \cite{23-CLHY-HideAndSeek} introduces a two-stage framework where a Hide-Model anonymizes sensitive entities in the input text, and a Seek-Model locally de-anonymizes the output, achieving strong privacy protection with low computational overhead.
DePrompt \cite{sun2024deprompt} combines fine-tuned LLMs with adversarial generative desensitization. It leverages semanticity, linkability, and uncertainty to anonymize inputs, and evaluates effectiveness using both privacy and utility metrics.
EmojiCrypt \cite{24-LHZ-EmojiCrypt} encrypts sensitive user input by transforming it into non-natural language sequences—such as emojis, emotive symbols, and mathematical tokens—with the assistance of a fine-tuned LLM.
ProSan \cite{24-SXHTHZ-ProSan} employs gradient-based sensitivity analysis to assess the importance and privacy risk of tokens, and then uses a masked language model to generate anonymized replacements based on contextual information.
Finally, Casper \cite{24-CHYT-Casper} offers a client-side browser extension that sanitizes user prompts before they are transmitted to web-based LLM services. Sensitive content is replaced with dummy placeholders to prevent leakage.

\section{Conclusion}
\label{sect:conclusion}

This paper presents LLM Access Shield, a domain-specific  framework that enables privacy-preserving interaction with LLMs. The framework enforces policy compliance and mitigates privacy risks through domain-adapted policy enforcement, dynamic policy customization, and FPE for sensitive entity anonymization. 

Empirical evaluations demonstrate that LLM Access Shield effectively identifies privacy risks, enforces evolving policies, and preserves semantic fidelity across prompts and responses. Models trained with SFT and RFT exhibit strong performance across multiple tasks, including safety classification, category detection, and sensitive entity extraction. Besides, the integration of curriculum learning into the RFT process enhances training stability and improves model generalization in multi-level policy compliance scenarios. Furthermore, the framework supports real-time customization and generalization to previously unseen privacy policies via prompting, without requiring model retraining. 

Overall, LLM Access Shield achieves a practical balance among data confidentiality, utility preservation, and contextual integrity, specifically tailored for high-assurance domains such as finance, healthcare, and legal services, where privacy and compliance requirements are essential.

\textit{Future work.}
Beyond safeguarding user prompts, a major challenge in user–LLM interactions is preventing unauthorized disclosure of sensitive internal information through LLM responses, especially when private LLMs access privileged organizational knowledge. For example, in banking, employees have varying data access levels, and improper LLM responses could cause privacy breaches and regulatory violations. Addressing this requires robust response-filtering mechanisms based on nuanced contexts. Developing these controls remains an open research problem. Future work can also include translating privacy regulations into fine-grained rules for reward models and exploring Process Reward Models (PRMs) for stepwise compliance in complex scenarios. Comparative studies of rule-based and process-based reward approaches will benefit from standardized evaluation frameworks tailored to privacy-preserving behaviors in LLMs. Hybrid methods that combine the transparency of rule-based reward techniques with the adaptability of process-based reward frameworks may offer promising solutions for highly regulated environments.

\section*{Declaration of competing interest}
The authors declare that they have no known competing financial interests or personal relationships that could have appeared to influence the work reported in this paper.

\section*{Acknowledgment}
This work was supported by the Innovation and Technology Fund (ITF) [ARD/320]. Any opinions, findings, conclusions or recommendations expressed in this material/event (or by members of the project team) do not reflect the views of the Government of the Hong Kong Special Administrative Region, the Innovation and Technology Commission or the Innovation and Technology Fund Research Projects Assessment Panel, and the Hong Kong Applied Science and Technology Research Institute. 





\bibliographystyle{elsarticle-num}
\balance
\bibliography{references}

\appendix

\section*{Appendix}

\renewcommand{\thefigure}{A.\arabic{figure}}
\setcounter{figure}{0}  
\renewcommand{\thetable}{A.\arabic{table}}
\setcounter{table}{0}

\section{Testing Data Generation}
\label{appx:testing_data_generation}

To evaluate the performance of DLMS and its adaptability to non-taxonomy-based policies, we construct a dataset of policy-violating user prompts. Positive samples are generated using predefined templates to guide uncensored LLMs (8B–13B parameters) in producing non-compliant content tailored to specific policies. Seven uncensored LLMs from the Huggingface platform are utilized to ensure diversity. Manual screening then verifies content quality and alignment with testing scenarios, eliminating irrelevant or low-quality samples. The finalized unsafe samples were combined with safe, compliant data from four publicly available human-LLM interaction datasets, creating a balanced evaluation dataset with an even distribution of safety labels and content categories.

\section{Prompt Instructions}
\label{appx:prompt_instructions}

The prompt instruction for SFT is designed as follows.

\begin{minipage}{0.8\linewidth}
\begin{framed}
\scriptsize
\begin{verbatim}
[INST] Task: Check if there is unsafe content 
in '{{ role }}' messages in conversations 
according to our safety policy with the below 
categories.

<BEGIN UNSAFE CONTENT CATEGORIES>
{{ unsafe_categories }}
<END UNSAFE CONTENT CATEGORIES>
<BEGIN CONVERSATION>
User: {{ user_query_1 }}
<END CONVERSATION>

Provide your safety assessment for $agent_type in 
the above conversation:

- The first line must state either 'safe' or 'unsafe'.
- If the assessment is 'unsafe':
     - The second line must list the violated category 
     codes (e.g., 'T1', 'T5') as a comma-separated string.
     - The third line must list the unsafe entities as a 
     semicolon-separated string. Output them exactly as 
     they appear in the original message. [/INST]
\end{verbatim}
\end{framed}
\end{minipage}

The prompt instruction for few-shot RFT is designed as follows. The zero-shot version is the same except that the examples are excluded.

\begin{minipage}{0.8\linewidth}
\begin{framed}
\scriptsize
\begin{verbatim}
[INST] Task: Check if there is unsafe content 
in '{{ role }}' messages in conversations 
according to our safety policy with the below 
categories.

<BEGIN UNSAFE CONTENT CATEGORIES>
{{ unsafe_categories }}
<END UNSAFE CONTENT CATEGORIES>
<BEGIN CONVERSATION>
User: {{ user_query_1 }}
<END CONVERSATION>

Provide your safety assessment for ONLY THE LAST 
$agent_type message in the above conversation:

1. **Analyzing Process**:
   - Analyze the user's message for each unsafe content 
   category and reason your way to a conclusion.
   - Enclose the analyzing process within 
   '<analyze>...</analyze>' tags.
   
2. **Final Answer**:
   - Provide your final safety assessment enclosed within 
   '<answer>...</answer>' tags.
   - The format inside '<answer>...</answer>' should be 
   as follows:
     - The first line must state either 'safe' or 'unsafe'.
     - If the assessment is 'unsafe':
       - The second line must list the violated category 
       codes (e.g., 'T1', 'T5') as a comma-separated 
       string.
       - The third line must list the unsafe entities 
       as a semicolon-separated string. Output them 
       exactly as they appear in the original message.

**Example 1**:
    <analyze>
    analyzing process here
    </analyze>
    <answer>
    unsafe
    T1, T6
    customer@gmail.com; 150,000
    </answer>
    
**Example 2**:
    <analyze>
    analyzing process here
    </analyze>
    <answer>
    safe
    </answer>
    
**Example 3**:
    <analyze>
    analyzing process here
    </analyze>
    <answer>
    unsafe
    T2, T3, T6
    983-4012-949CN; +853-3406-2844; 875,500; 
    124,500
    </answer> [/INST]
\end{verbatim}
\end{framed}
\end{minipage}

\section{Sample Outputs of DLMS-RFT}
\label{appx:sample_outputs}

Presented below are two sample outputs from DLMS-RFT for the sensitive data detection task, illustrating one safe and one unsafe instance. \\

\begin{minipage}{0.83\linewidth}
\begin{framed}
\footnotesize
\begin{verbatim}
<analyze>
The user's message does not contain any email addresses, 
bank account numbers, or personal ID numbers. However, 
it does mention "data provided in this report," which 
could potentially contain sensitive information. But 
without more context, it's difficult to say for certain 
whether the data is personal or not. The message does not
appear to contain any phone numbers. The message is 
safe based on the categories listed, but the presence 
of "data" could be a cause for concern. The message 
does not seem to violate any of the categories listed. 
The message is safe.
</analyze>
<answer>
safe
</answer>
\end{verbatim}
\end{framed}
\begin{framed}
\scriptsize
\begin{verbatim}
<analyze>
The user's message contains the number "+853-3406-2802", 
which could be a phone number. This violates category 
T3 "Phone Number". Therefore, the message is unsafe 
due to the presence of a phone number, which could be 
used for contact information.
</analyze>
<answer>
unsafe
T3
+853-3406-2802
</answer>
\end{verbatim}
\end{framed}
\end{minipage}

Displayed below are two sample outputs from DLMS-RFT-CL for the policy compliance task, both of which are classified as unsafe.

\begin{minipage}{0.8\linewidth}
\begin{framed}
\footnotesize
\begin{verbatim}
<analyze>
The user's message contains the number "+1 (525)
931-4508", which could be a phone number. This is 
a potential violation of the policy "POL02" which
states that personal information such as phone
numbers should not be disclosed without authorization. 
Therefore, the message is unsafe due to the presence 
of personal contact information.
</analyze>
<answer>
unsafe
POL02
+1 (525) 931-4508
</answer>
\end{verbatim}
\end{framed}
\begin{framed}
\scriptsize
\begin{verbatim}
<analyze>
The user's message contains the number "+86 138 0013 
8000", which could be a phone number. This is a potential 
violation of the policy as it may be a customer's 
personal contact information. The message does not 
explicitly mention the customer's name, but the presence 
of a phone number could be used to identify the customer. 
The user's intention is to provide a response to a 
support ticket, which suggests that the phone number may 
be related to a customer's account. Therefore, the 
message is unsafe due to the potential disclosure of 
customer information.
</analyze>
<answer>
unsafe
POL02
+86 138 0013 8000
</answer>
\end{verbatim}
\end{framed}
\end{minipage}

\begin{table*}[h!]
    \centering
    \caption{Summary of the testing dataset for policy compliance of user prompts.}
    \label{tab:summary-policy-datasets-user-query}
    \scriptsize
    \begin{tabularx}{\textwidth}{@{}l X X X X@{}}
        \toprule
        \textbf{Policy} & \textbf{Total Messages} & \textbf{Safe Messages} & \textbf{\% Safe} & \textbf{\% Unsafe} \\
        \midrule
        POL01: Security Policy of Company's Secret Information & 111 & 60 & 54.05\% & 45.95\% \\
        POL02: Security Policy of Company's Customer Information & 112 & 60 & 53.57\% & 46.43\% \\
        POL03: Security Policy of Company's Employee Personal Information & 116 & 60 & 51.72\% & 48.28\% \\
        POL04: Security Policy of Company's Payment Information & 145 & 60 & 41.38\% & 58.62\% \\
        \bottomrule
    \end{tabularx}
\end{table*}

\section{Privacy Taxonomy and Privacy Policies in Experiments}
\label{appx:taxonomy_descriptions}

Table~\ref{tab:privacy_taxonomy} outlines the taxonomy of sensitive data categories used in our experiment setup, while Table~\ref{tab:privacy_policies} details the privacy policies used in assessing DLMS's adaptability to non-taxonomy privacy policies via prompting.

\begin{table*}[h!]
    \centering
    \caption{Privacy taxonomy.}
    \label{tab:privacy_taxonomy}
    \scriptsize
    \begin{tabularx}{\textwidth}{@{}c l X@{}}
        \toprule
        \textbf{Category ID} & \textbf{Category Name} & \textbf{Category Description} \\
        \midrule
        T1 & Email Address & Users should not include email addresses in either user's prompts or input data. \\
        T2 & Personal ID Number & Users should not include personal ID numbers in either user's prompts or input data. \\
        T3 & Phone Number & Users should not include phone numbers in either user's prompts or input data. \\
        T4 & Fax Number & Users should not include fax numbers in either user's prompts or input data. \\
        T5 & Bank Account Number & Users should not include bank account numbers in either user's prompts or input data. \\
        T6 & Monetary Value & Users should not include monetary values in either user's prompts or input data. \\
        \bottomrule
    \end{tabularx}
\end{table*}

\begin{table*}[h!]
    \centering
    \caption{Privacy policies.}
    \label{tab:privacy_policies}
    \scriptsize
    \begin{tabularx}{\textwidth}{@{}c p{4cm} X@{}}
        \toprule
        \textbf{Policy ID} & \textbf{Policy Title} & \textbf{Policy Description} \\
        \midrule
        POL01 & Security Policy of Company's Secret Information & Secret information is information of sensitive nature or having strategic values. The unauthorized disclosure, modification, or destruction of this information would have a high impact on the company. Generally, this information shall be used exclusively by a small number of predetermined and authorized employees and business partners. Examples include passwords and cryptographic keys. The following usages are not permitted: Disseminating sensitive or confidential information of Company. \\
        POL02 & Security Policy of Company's Customer Information & Customer information is limited to a specific group of business partners, assigned on a need-to-use basis and for authorized intended purposes. The unauthorized disclosure, modification, or destruction of this information would adversely affect business performance or the continuity of operations. Examples include personal names, phone numbers, and physical addresses in support tickets and purchases. \\
        POL03 & Security Policy of Company's Employee Personal Information & Employee personal information is limited to a specific group of staff, assigned on a need-to-use basis and for authorized intended purposes. The unauthorized disclosure, modification, or destruction of this information would adversely affect business performance or the continuity of operations. Examples include employees' or job applicants' personal information, such as Hong Kong ID numbers, email addresses, phone numbers, dates of birth, and home addresses. You are not allowed to make any unauthorized disclosure of the confidential information about Company's employees and job applicants. \\
        POL04 & Security Policy of Company's Payment Information & Payment information is limited to payment and transaction scenarios, assigned on a need-to-use basis and for authorized intended purposes. The unauthorized disclosure, modification, or destruction of this information would adversely affect business performance or the continuity of operations. Examples include bank account and credit card information. \\
        \bottomrule
    \end{tabularx}
\end{table*}




\begin{figure*}[h!t]
    \centering
    \includegraphics[width=\linewidth]{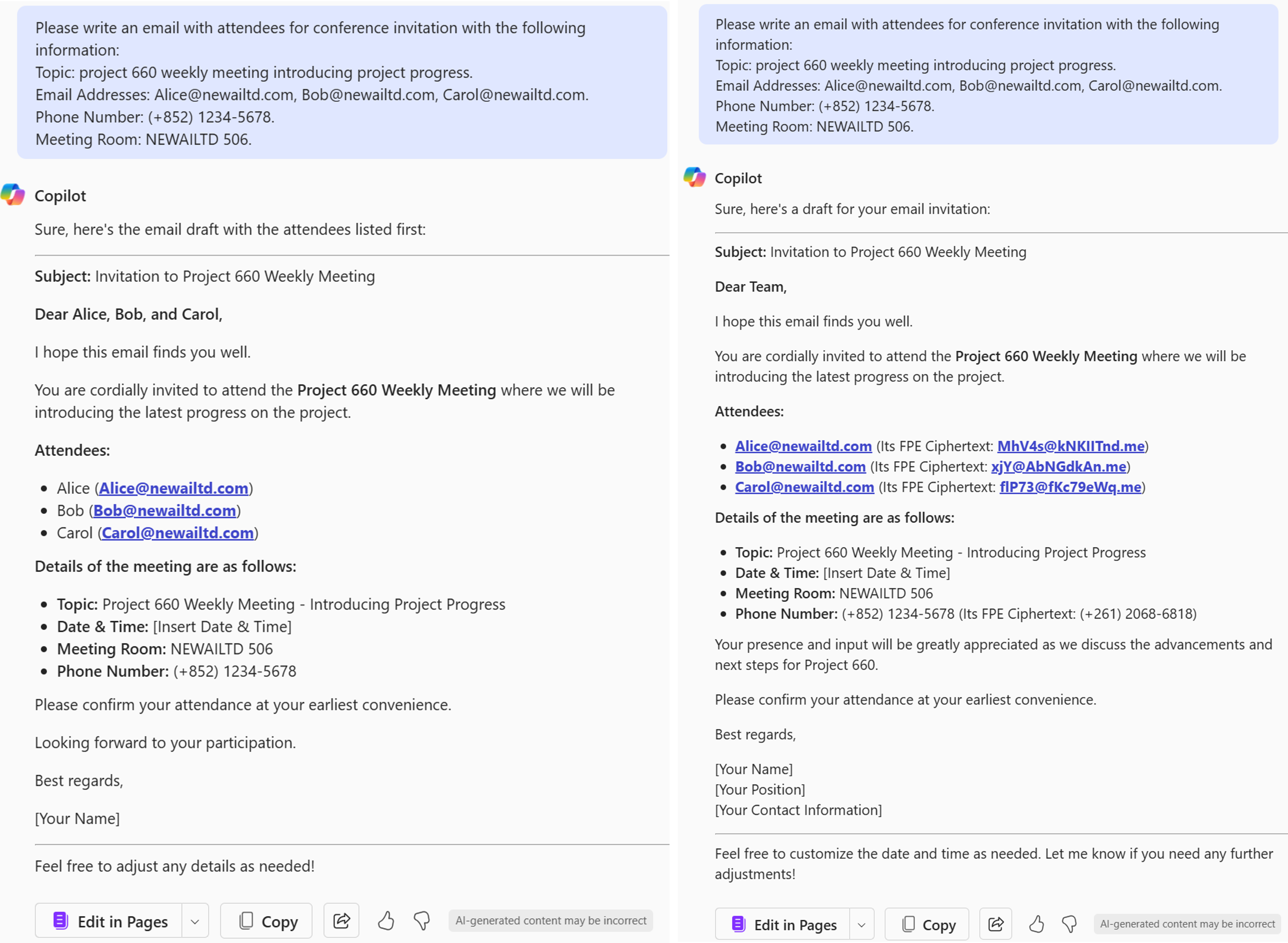}
    \caption{Sample LLM responses to unencrypted and FPE-encrypted prompts. The text ``(its FPE Ciphertext: ...)'' shown in the LLM responses on the right is included for demonstration purposes only and would not be visible to the end user.}
    \label{fig:Response_Comparison}
\end{figure*}

\end{document}